\documentclass{jkas}
\include{graphicsx}

\def\year{2019} 
\def\month{September} 
\def\volume{00} 
\def\issue{0} 
\def\beginpage{1} 
\def\endpage{99} 
\def\received{---} 
\def\accepted{---} 

\date{Received \received ; accepted \accepted}

\title{
Osculating Versus Intersecting Circles in Space-Based Microlens Parallax Degeneracies
}

\author[1,2]{A.~Gould}

\affil[1]{Max-Planck-Institute for Astronomy, K\"{o}nigstuhl 17, 69117 Heidelberg, Germany}
\affil[2]{Department of Astronomy Ohio State University,
140 W.\ 18th Ave., Columbus, OH 43210, USA 
\email{gould@astronomy.ohio-state.edu }}

\def\corrauthor{
Andrew Gould
}

\def\runningauthor{
Andrew Gould
}

\def\runningtitle{
Osculating Versus Intersecting Circles
}

\def\keywords{
gravitational microlensing  
}

\def\abstracttext{
I investigate the origin of arc degeneracies in satellite microlens
parallax $\bpi_\e$ measurements with only late time data, e.g., $t>t_0
+ t_\e$ as seen from the satellite.  I show that these are due to
partial overlap of a series of osculating, exactly circular,
degeneracies in the $\bpi_\e$ plane, each from a single measurement.
In events with somewhat earlier data, these long arcs break up into
two arclets, or (with even earlier data) two points, because these
earlier measurements give rise to intersecting rather than osculating
circles.  The two arclets (or points) then constitute one pair of
degeneracies in the well-known four-fold degeneracy of space-based
microlens parallax.  Using this framework of intersecting circles, I
show that next-generation microlens satellite experiments could yield
good $\bpi_\e$ determinations with only about five measurements per
event, i.e., about 30 observations per day to monitor 1500 events per
year.  This could plausibly be done with a small (hence cheap, in the
spirit of Gould \& Yee 2012) satellite telescope, e.g., 20 cm.
}

\newcommand{\bdv}[1]{\mbox{\boldmath$#1$}}

\def\au{{\rm AU}} 
 
\def\bv{{\bf v}} 
\def\bQ{{\bf Q}}

\def\mas{{\rm mas}}

\def\rel{{\rm rel}}

\def\eff{{\rm eff}}
\def\sat{{\rm sat}}
\def\hel{{\rm hel}}
\def\geo{{\rm geo}}

\def\lim{{\rm lim}}
\def\e{{\rm E}}
\def\bpi{{\bdv\pi}}
\def\bmu{{\bdv\mu}}

\def\btheta{{\bdv\theta}}

\def\bD{{\bf D}}
\def\bP{{\bf P}}

\def\apj{{ApJ}}
\def\aj{{AJ}}
\def\apjl{{ApJL}}

\def\aap{{A\&A}}

\def\mnras{{MNRAS}}

\begin{document}
\jkashead 


\section{{Introduction}
\label{sec:intro}}

In his original paper on space-based microlens parallax measurements,
\citet{refsdal66} already noted that they were subject to a discrete 
four-fold degeneracy.  Two observatories, one on Earth and one on a satellite,
would each see a single-lens single-source (1L1S) microlensing event,
characterized by three \citet{pac86} parameters $(t_0,u_0,t_\e)$,
but these parameters would differ due to their different viewpoints.
Here $t_0$ is the time of maximum magnification, $u_0$ is the impact parameter
normalized to the Einstein radius $\theta_\e$, 
and $t_\e$ is the Einstein timescale,
\begin{equation}
t_\e\equiv {\theta_\e\over\mu_\geo};
\quad
\theta_\e^2 = \kappa M \pi_\rel,
\label{eqn:tedef}
\end{equation}
where $M$ is the mass of the lens, $(\pi_\rel,\bmu_\geo)$ are the
lens-source relative (parallax, proper motion) and 
$\kappa\equiv 4 G/c^2\au\simeq 8.14\,\mas\,M_\odot^{-1}$.  In more
modern language \citep{gould00,gould04,gouldhorne}, the microlens
parallax vector,
\begin{equation}
\bpi_\e \equiv {\pi_\rel\over\theta_\e}\,{\bmu_\geo\over\mu_\geo},
\label{eqn:piedef}
\end{equation}
could be determined from the inferred offset in the Einstein ring
\begin{equation}
\bpi_\e = {\au\over D_\perp}(\Delta\tau^\prime,\Delta\beta^\prime);
\label{eqn:pieeval}
\end{equation}
where
\begin{equation}
\Delta\tau^\prime = {t_{0,\sat} - t_{0,\oplus}\over t_\e};
\qquad
\Delta\beta^\prime = {u_{0,\sat} - u_{0,\oplus}},
\label{eqn:dtaudbeta}
\end{equation}
and $\bD_\perp$ is the two dimensional (2-D) vector offset from Earth
to the satellite projected on the sky (approximated as a constant during
the observations).  The first component is then
along this direction and the second is perpendicular to it.
The four-fold degeneracy arises from the fact that only the magnitude
(but not the sign) of $u_0$ can generally be inferred from the light curve.
See Figure~1 from \citet{gould94}.

The great majority of subsequent theoretical work on space-based microlens
parallax (and it degeneracies) took place within the context of events
for which there were reasonably complete light-curve measurements from
both Earth and the satellite, so that in particular it was possible
to measure $(t_0,u_0)_\sat$.  For example, while \citet{refsdal66} had
suggested observations from a second satellite to break the four-fold
degeneracy, \citet{gould95} argued that this might be possible from
a single satellite because the velocity difference between the
two observatories would yield differences in $t_\e$ that would allow
one to distinguish among the four values of 
$\Delta\beta^\prime_{\pm,\pm}$,
where the first subscript refers to the sign of $u_{0,\oplus}$ and
the second to $u_{0,\sat}$.  This was soon shown to be substantially
more efficient for microlensing events toward the ecliptic poles
\citep{boutreux96} than toward the ecliptic \citep{gaudi97}.

A key issue in these early years appeared to be the much greater
difficulty in measuring $u_0$ compared to $t_0$ for 1L1S light curves.
This arises from the fact that the derivative of the microlensed flux
with respect to only one parameter ($t_0$) is odd (antisymmetric) in time,
while there are four with derivatives that are even (symmetric) in time
($u_0,t_\e,f_s,f_b$).  Here $(f_s,f_b)$ are the source flux and
blended flux.  Hence, $u_0$ is strongly correlated with other parameters
while $t_0$ is not.  \citet{gould95} already recognized that the interplay
of discrete and continuous degeneracies in the direction orthogonal
to $\bD_\perp$ was a major issue for space-based parallaxes because it
seemed to require very high signal-to-noise ratio space-based light curves,
which are intrinsically expensive.  He noted that if the space and
ground cameras had nearly identical responses, then this issue could
be largely resolved.  This is because $f_s$ would be known to be the same
a priori, which would allow $\Delta\beta^\prime = (u_{0,\sat}-u_{0,\oplus})$ to
be measured much more precisely than either impact parameter separately.
However, this was believed to be extremely difficult even for optical
observations and essentially impossible for the only photometric
telescope then planned for solar orbit, namely SIRTF (a.k.a., {\it Spitzer}),
whose shortest wavelength ($3.6\,\mu$m) was essentially unobservable
from the ground.

Pressed by M.\ Werner (1998, private communication) to find a solution
to this problem that could be applied to {\it Spitzer}, \citet{gould99}
developed the idea of combining separate one-dimensional (1-D) parallax information from Earth
and {\it Spitzer} to yield robust 2-D microlens parallaxes.  That is,
according to Equation~(\ref{eqn:pieeval}), the component of $\bpi_\e$
along $\bD_\perp$ could be well measured even if $u_{0,\sat}$ (and so
$\Delta\beta^\prime$) was not.
Therefore, if there were additional 1-D information from the ground
(not parallel to $\bD_\perp$), then a relative handful of space-based
measurements (enough to measure $t_{0,\sat}$) would be sufficient.

In fact, \citet{gmb94} had already pointed out that the annual parallax
effect \citep{gould92} could measure the component of $\bpi_\e$
parallel to Earth's instantaneous acceleration at $t_0$, even
when the orthogonal component was essentially 
unmeasurable\footnote{Subsequently \citet{smp03} studied this
much more deeply and showed that the parallel component is third order
in time while the perpendicular component is fourth order.}.  Thus,
unless Earth's acceleration at $t_0$ is closely aligned with $\bD_\perp$,
the two 1-D parallaxes (each by itself almost useless) could be combined
to yield a 2-D parallax.  This led to a proposal for target-of-opportunity
observations toward the Magellanic Clouds (where these two directions
are generally not aligned) and resulted in a successful measurement
based on just four {\it Spitzer} epochs \citep{os05001}.

The extremely high cost (hence low expected number) of space-based 
measurements led \citet{gouldyee12} to suggest a radically different
idea for ``cheap space-based microlens parallaxes''.  This required two special
conditions.  First, the event must be relatively high-magnification
as seen from Earth ($u_{0,\oplus}\ll 1$).  Second, it must be observed from
the satellite at a time $t_\sat \simeq t_{0,\oplus}$.  However, if these
two conditions could be met (and if there were an additional late-time
measurement to determine the baseline flux, $f_{\rm base,\sat}$), then one
could determine the flux difference 
$\Delta f_\sat = f_\sat(t_\sat)-f_{\rm base,\sat}$,
and thus the magnification $A_\sat$ and corresponding offset in the
Einstein ring $u_\sat$:
\begin{equation}
A_\sat = 1+{\Delta f_\sat\over f_{s,\sat}};
\qquad
u_\sat =\sqrt{2\Biggl({1\over\sqrt{1-A_\sat^{-2}}}-1\Biggr)}.
\label{eqn:asat}
\end{equation}
Then, in the approximation $u_{0,\oplus}\rightarrow 0$, the magnitude
of the parallax vector is simply $\pi_\e = (\au/D_\perp)u_\sat$.
There is then no information at all about the direction $(\phi_\pi)$
of $\bpi_\e$, but this direction is not needed to determine the
main properties of the lens, i.e., its mass $M=\theta_\e/\kappa\pi_\e$
and lens-source relative parallax $\pi_\rel=\theta_\e\pi_\e$.

Of course, this requires that $f_{s,\sat}$ be known, which in the
previous conception required a good-coverage, high-precision, space-based
light curve.  However, in the meantime, \citet{mb11293} had established
that microlensing source fluxes of sparsely covered light curves
could be determined from color-color relations linked to well covered
light curves.  Hence, \citet{gouldyee12} suggested that these relations 
be applied to space-based observations as well.

Subsequently, \citet{ob161045} demonstrated that this approach works
in practice.  In particular, their Figure~3, which shows a circle
nearly centered on the origin (excellent measurement of $\pi_\e$,
no information on $\phi_\pi$) was a major inspiration for the present
work.

For 2014-2019, there were (or will be) major {\it Spitzer} microlens
parallax campaigns toward the Galactic bulge.  During the first (pilot)
year, the focus was on obtaining ``full-coverage'' light curves
from {\it Spitzer}, in particular capturing the peak, in order
to demonstrate the feasibility of the method.  See, for example,
Figure~1 from \citet{ob140939} and compare to Figure~1 of \citet{gould94}.
However, in subsequent years, the criteria for event selection
were substantially relaxed in pursuit of the goal of measuring
the Galactic distribution of planets \citep{yee15b}.  In particular,
events were frequently chosen even if the {\it Spitzer} observations
were likely to begin well after peak.  As discussed above, such light-curve 
fragments cannot by themselves yield useful information about
$(t_0,u_0)_\sat$.  However, it was anticipated (and subsequently
confirmed, \citealt{170event}) that $f_{s,\sat}$ can be derived
from color-color relations (provided that $f_{s,\oplus}$ is well measured
from Earth).

Nevertheless, despite the fact that there are now several hundred
{\it Spitzer} light curves that begin after peak, there has not
yet been a systematic study of what is the character of the 
parallax information that is actually garnered from these light curves.
Rather, {\it Spitzer} and ground-based data are generally combined
in a single fit, often after considering models based on
ground-based data alone.  However, an important exception to this
approach was taken by \citet{ob180596}.  Their ``{\it Spitzer}-only''
parallax contours (Figure~5, left panels) look very much like arcs
of a circle, but in contrast to circles of \citet{ob161045}, they
are not centered on the origin.  This suggests that the parallax information
content of late-time satellite light curves may be intrinsically
circular.  If so, a deeper understanding of the origin of this
effect will be valuable for both planning and interpreting microlensing
parallax observations.  I therefore undertake such an investigation
here.

\section{{Idealized Case: Single Observation at Late-time Epoch}
\label{sec:ideal}}

Let us consider a high-magnification (i.e., $u_{0,\oplus}\ll 1$) 
microlensing event, with peak time $t_{0,\oplus}$
as seen from Earth.  And let us assume that there are two late-time
measurements from a satellite, one at $t_\sat$ and the other at baseline.
As discussed in Section~\ref{sec:intro},
given a color-color relation, this leads via Equation~(\ref{eqn:asat})
to successive determinations of $\Delta f_\sat$,
$A_\sat$, and $u_\sat(A_\sat)$.

I follow \citet{calchi16} in working within a heliocentric
framework, but present the results in geocentric quantities, in particular
the parallax $\bpi_\e$ and Einstein timescale, $t_\e$, from which I omit
the ``geocentric subscripts''.
The geocentric and heliocentric projected velocities are 
given by 
\begin{equation}
\tilde \bv = {\bpi_\e\over\pi_\e^2}{\au\over t_\e}; 
\qquad 
\tilde \bv_\hel = \tilde \bv + \bv_{\oplus} ,
\label{eqn:tidlev}
\end{equation}
where $\bv_{\oplus}$ is the 2-D vector representing the 
instantaneous motion of Earth relative to the Sun at $t_{0,\oplus}$ projected
on the sky.

Let $\bD$ be the 2-D separation vector (again projected on the sky)
between Earth's position at $t_{0,\oplus}$ and the satellite's position 
at $t_\sat$.  (Notice that this is different than the definition of $\bD_\perp$ 
given in Section~\ref{sec:intro}, and for this reason I use a different symbol.)
And let
\begin{equation}
\Delta t = t_\sat - t_0;
\qquad
\Delta\tau = {\Delta t\over t_\e}.
\label{eqn:ddtdef}
\end{equation}

Then,
\begin{equation}
u_\sat^2 = \bigg|{\tilde\bv_\hel\Delta t -\bD\over \tilde r_\e}\bigg|^2
= \bigg|{\tilde\bv\Delta t-(\bD-\bv_\oplus\Delta t)\over\au/\pi_\e}\bigg|^2,
\label{eqn:usat0}
\end{equation}
\begin{equation}
u_\sat^2= \bigg|{\bpi_\e\Delta\tau\over \pi_\e}-\bQ\pi_\e\bigg|^2,
\label{eqn:usat1}
\end{equation}
where $\tilde r_\e \equiv\au/\pi_\e$ is the projected Einstein radius in the
observer plane
and
\begin{equation}
\bQ\equiv {\bD - \bv_\oplus\Delta t\over \au}.
\label{eqn:qdef}
\end{equation}
That is,
\begin{equation}
u_\sat^2 = (\Delta\tau)^2 -2\bQ\cdot\bpi_\e\Delta\tau + Q^2\pi_\e^2
= \bigg|Q\bpi_\e - {\bQ\Delta\tau\over Q}\bigg|^2,
\label{eqn:usat2}
\end{equation}
or
\begin{equation}
\biggl(\bpi_\e-{\bQ\Delta\tau\over Q^2}\biggr)^2 =\biggl({u_\sat\over Q}\biggr)^2.
\label{eqn:circle}
\end{equation}
Hence, such a single-epoch space-based observation yields a 
circular $\bpi_\e$ contour of radius
$u_\sat/Q$ and center $\bQ\Delta\tau/Q^2$.  The solution to 
Equation~(\ref{eqn:circle}) can be written in parametrized form
\begin{equation}
\bpi_\e ={\bQ\Delta\tau\over Q^2} + {u_\sat\over Q}\hat{\bf n}
\label{eqn:circle2}
\end{equation}
where $\hat{\bf n}$ represents a unit vector in an arbitrary direction.

\begin{figure}
\centering
\includegraphics[width=90mm]{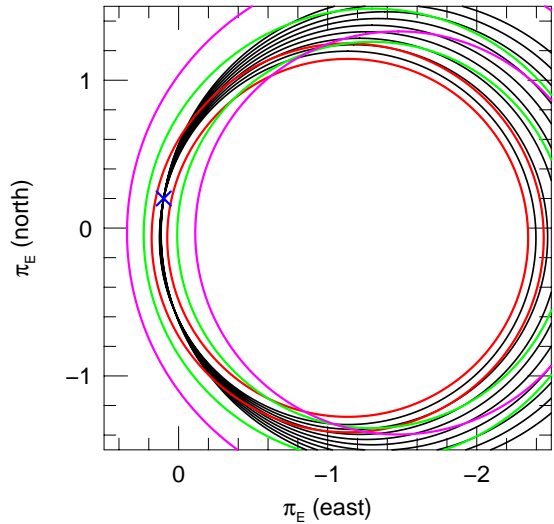}
\caption{Evolution of parallax circles from individual
photometric measurements.  The black circles are the locus of
$\bpi_\e$ consistent with individual measurements 
after (1,5,9,13,17,21,25,29,33) days of daily {\it Spitzer} observations,
under the assumption of perfect ($\sigma = 0$) photometric measurements,
for a hypothetical $t_\e=30\,$day event that peaks at $u_{0,\oplus}=0$ on
$t_0=\,$25 May 2019 and 
for which the first {\it Spitzer} observation is 9 July 2019, i.e.,
at $\Delta\tau=1.5$.  The parallax is $(\pi_{\e,N},\pi_{\e,E})=(0.2,0.1)$
(blue cross).
The pairs of red, green, and magenta circles show the $1\,\sigma$
error range for the measurements at days 1, 17, and 33, respectively, assuming
photometric measurement errors $\sigma = 0.01\,$mag.  While the black
circles all cross the true parallax value, the finite $1\,\sigma$ ranges
(which grow with time) lead to a joint solution in the shape of an arc.
See Figure~\ref{fig:all_0}.
}
\label{fig:circ}
\end{figure}

I note that for simplicity of exposition, I have imagined satellite
observations that take place well after Earth-based peak, i.e.,
$\Delta t = t_\sat - t_{0,\oplus}>0$.  However, the formula applies
equally well to single observations that are taken {\it at any time}.
In particular, this includes single observations that take
place well before Earth-based peak,
i.e., $\Delta t <0$.  There are many practical cases of this
in real observations as well.

\section{{Impact of Realistic Conditions on Ideal Case}
\label{sec:impact}}

Equation~(\ref{eqn:circle}) applies quite generally to the idealized 
case.  However, because almost 800 microlensing events have been observed 
with {\it Spitzer}, it is important to understand how this idealization
relates to this ensemble of real observations.

\begin{figure}
\centering
\includegraphics[width=90mm]{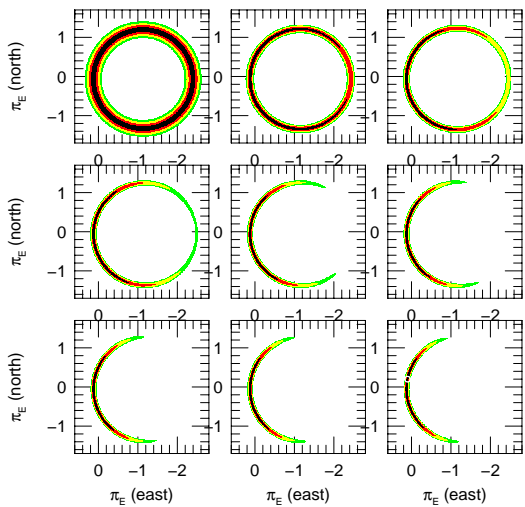}
\caption{Evolution of an arc.  Successive panels show 
the $\bpi_\e$ error contours
after (1,5,9,13,17,21,25,29,33) days of daily {\it Spitzer} observations,
with measurement errors $\sigma = 0.01\,$mag,
for the same hypothetical event illustrated in Figure~\ref{fig:circ}, which
has true parallax $(\pi_{\e,N},\pi_{\e,E})=(0.2,0.1)$.
The colors (black, red, yellow, green) indicate $\Delta\chi^2 < (1,4,9,16)$.
The contours evolve slowly from a circle to an arc, but cease after the
sixth panel $(\Delta\tau=2.4)$ because the errors on the radii ($u_\sat/Q)$ 
of the successive degenerate circles 
become too large for the additional observations
to contribute.  See Equation~(\ref{eqn:usaterr0}) and Figure~\ref{fig:circ}.
}
\label{fig:all_0}
\end{figure}

One important practical point to keep in mind is that for {\it Spitzer}
observations toward the bulge, the vector $\bQ$ points roughly due west and
its amplitude lies approximately in the range
\begin{equation}
Q \simeq 1 +\sin[8^\circ\times(Y - 2013) \pm 18^\circ],
\label{eqn:qeval}
\end{equation}
where $Y$ is the year of observation.  The direction simply
reflects the facts that {\it Spitzer} is in an Earth-trailing orbit and
that for Galactic-bulge targets, the ecliptic is roughly parallel
to the equator.  Then, because the 2014-2019 campaigns have taken
place when the bulge is approximately in opposition
(while $t_{0,\oplus}$ is almost always within $\sim 1.5\,$ months of
opposition), the $\bv_\oplus\Delta t$
term in Equation~(\ref{eqn:qdef}) approximately ``corrects'' the
Earth position going into ``$\bD$'' to what it would be
at the time of the {\it Spitzer} observation.  On the other hand,
due to Sun-angle restrictions, {\it Spitzer} observations toward
the ecliptic are always near quadrature.  This accounts for the
form of Equation~(\ref{eqn:qeval}).  The normalization and range
reflect the fact that in 2013, the bulge was in opposition at
the midpoint of the 38-day {\it Spitzer} viewing window.

Continuing to restrict attention to ``high-magnification'' ($u_{0,\oplus}\ll 1)$
events, there are two main differences between real observations
and the idealized case of Section~\ref{sec:ideal}.  First, there
are in practice not just two observations, but a series of observations
that either begin well after $t_{0,\oplus}$ or end well before 
$t_{0,\oplus}$.  Second, the value of $u_\sat$ for each observation 
is not known precisely but with some finite error.  

Regarding the errors,
both quantities that enter the circle center 
in Equation~(\ref{eqn:circle2}) ($\bQ$ and $\Delta\tau$) 
are precisely known, so the only
uncertainty in the description of the circle is in its radius.
This derives from the error in the value of $u_\sat$,
which propagates via Equation~(\ref{eqn:asat}) from
$A_\sat = 1 + (f_\sat - f_{\rm base,\sat})/f_{s,\sat}$.  Thus, there
are three potential sources of error: the individual measurement
error $f_\sat$, the estimate of the baseline flux $f_{\rm base,\sat}$,
which in practice comes from the overall fit to the satellite light curve,
and  the satellite source flux $f_{s,\sat}$, which comes from the
color-color relation.  

The last of these puts a fundamental limit on the precision in the
sense that this error cannot be improved by additional observations.
However, as I now show, its impact is usually small.
The color-color relation yields an error in magnitudes, 
e.g., $\sigma=0.04\,$mag\footnote{Such errors reflect two steps: measuring
the source color in two bands from the ground and measuring a color-color
relation by cross-matching three-band space and ground photometry.
Both steps may require special efforts.  For example, ground-based
surveys routinely take sparse $V$-band observations to yield $V-I$
source colors, but {\it Spitzer} (at $L=3.6\,\mu$m) often observes
highly extincted targets for which the $V$ observations are practically
useless.  It is then essential to observe in a near-IR band (such as $H$)
while the event is substantially magnified to obtain a ground color
(e.g., \citealt{kb180029}).  With well-magnified data in two bands,
the ground color measurement is usually accurate to a few hundredths of
a magnitude.  It is also usually straightforward to obtain a precise
color-color relation of bulge stars (so, suffering similar extinction
to the source), but this is almost always restricted to giant stars,
whereas the sources are often dwarfs.  Depending on the source color
and the three photometric bands, giants and dwarfs can obey different 
color-color relations, and this must be carefully taken into account
(e.g., \citealt{ob161195}).  In general, errors of order this example
value are readily achieved provided that timely ground-based color
data are taken, but careful treatment is required.
}.  Propagating through Equation~(\ref{eqn:asat}),
we obtain $\sigma(A_\sat)= (A_\sat-1)k\sigma$, where $k=0.4\ln 10$, and so
\begin{equation}
\sigma_0(u_\sat) = {\sigma(A_\sat)\over |d A/du|} = 
\biggl({A-1\over A}\,{u(u^2+2)(u^2+4)\over 8}\biggr)k\sigma.
\label{eqn:usaterr0}
\end{equation}
The coefficient in brackets is relatively small and stable over the
relevant range of $u$, taking on values of $(0.29,0.22,0.27,0.37,0.50)$
for $u=(0.5,1.0,1.5,2.0,2.5)$.  Therefore, we expect that the limit on the 
width of the circle in the $\bpi_\e$ plane due to the color-color relation
will be small,
For example, for $\sigma=0.04\,$mag and $Q=1.3$, this limit would be
$\sigma_0 \lesssim 0.01$ over the range $0.5<u_\sat<2$.

The error due to the individual flux measurement errors (expressed
in magnitudes $\sigma_i$) degrades much more rapidly with increasing $u$.
Ignoring the other two sources of error (color-color relation and baseline
flux) and considering the case of zero blending, this can be evaluated
\begin{equation}
\sigma_i(u_\sat) = \biggl({u(u^2+2)(u^2+4)\over 8}\biggr)k\sigma_i.
\label{eqn:usaterr1}
\end{equation}
For the same five values of $u=(0.5,1.0,1.5,2.0,2.5)$, the
coefficient in brackets takes on
values of $(0.60,1.9,5.0,12,39)$.  Thus, for observations that begin
outside the Einstein ring $(u_\sat>1)$, the parallax information content
is dominated by the earlier observations.  This has important implications,
which I discuss immediately below.
The last source of error (in $f_{s,\rm base}$) generally plays the role
of exacerbating this effect: it is subdominant in the early observations,
while the later-time observations mainly contribute to evaluating
$f_{s,\rm base}$ itself.

Based on this assessment of the errors, I now show that 
the main impact of a finite series of observations (relative to a
single observation) is usually
to partially break the complete-circle degeneracy and turn it into an arc
(e.g., Figure~5 of \citealt{ob180596}).  The first point to note
is that for most late-starting observations, $Q\gg \pi_\e$.  That is,
typically $Q\sim 1$ while $\pi_\e\lesssim 0.2$ for most 
events\footnote{This limit applies to the great majority of bulge events
because $\pi_\rel\lesssim 0.03\,\mas$, while most lenses have masses 
$M\gtrsim 0.1\,M_\odot$.  Many disk lenses have $\pi_\e\lesssim 0.2$
as well, e.g., those lying more than halfway to the Galactic center
$(\pi_\rel<0.125\,\mas)$ with masses $M>0.4\,M_\odot$.}.  
Moreover the direction
of $\bQ$ changes very little with time because $\bD$ and $\bv_\oplus$
are both approximately aligned with the ecliptic.  Therefore,
the center of the circle $(\Delta\tau/Q)(\bQ/Q)$ is gradually moving
west while its eastern limb must always pass (within errors) through $\bpi_\e$,
which is near the origin.  Thus, the arcs comprising the eastern limbs
of circles from multiple epochs will largely coincide, while the western
limbs will increasingly separate, i.e., be inconsistent with one another.
See Figure~\ref{fig:circ}.
However, as discussed in the previous paragraph, the width of these
circles is rapidly increasing, so that most of their constraining
power comes from the earlier measurements.  For this reason, the
process tends to leave parallax arcs, which (other things being equal)
are longer for observations sequences that start at higher $u_\sat$.
See Figure~\ref{fig:all_0}.

\section{{Resolution of 1-D Degeneracy}
\label{sec:resolution}}

As discussed in Section~\ref{sec:impact}, multiple late-time measurements
will always restrict the circle described by Equations~(\ref{eqn:circle})
and (\ref{eqn:circle2}) to an arc.  And if these observations 
begin early enough,
then the arc (or arcs, see below) will be sufficiently restricted
to regard them as 2-D (rather than 1-D) measurements.  In fact,
if the measurements begin sufficiently early, one should just recover
the two-fold degeneracy\footnote{The degeneracy is only two-fold,
rather than four-fold, because we are still working in the regime
where $u_{0,\oplus}\sim 0$.} predicted by \citet{refsdal66} and illustrated
by Figure~1 of \citet{gould94}.  That is, with improving information,
the arc should break up into two arclets placed symmetrically
with respect to the $\bQ$ (essentially, $\bD$) axis.

However, in this section, I want to focus on how this 1-D arc
(or even circle) degeneracy can be broken for the cases that
the arc is relatively long.  There are two classes of methods: information from
annual parallax, and independent information about the direction of 
the lens-source relative proper motion $\bmu_\geo$.  For the second class,
there are three known distinct approaches.

\subsection{{Combining with 1-D Annual Parallax Measurements}
\label{lab:1Dpar}}

A very large fraction of microlensing events, at least among
those that are bright enough to allow {\it Spitzer} observations,
have sufficient information for 1-D parallax measurements.
These are usually straight in the Cartesian $\bpi_\e$ plane.
See, for example, 
Figure~3 of \citet{mb0337}, 
Figure~4 of \citet{ob03175}, 
Figure~2 of \citet{ob03238}, and 
Figure~1 of \citet{poindexter05}.  
The reason that these are all very old papers, from an era when
the rate of microlensing-event discovery was $\sim 5$ times lower
than today, is that the main scientific interest was in the effect
itself and its potential applications, rather than in the $\pi_\e$ measurement,
which was generally too weak to be useful.  However, there
have been some cases for which such 1-D measurements did play a
significant role in the immediate scientific results, e.g.,
Figure~2 of \citet{ob05071b}, 
Figure~3 of \citet{ob07050}, and
Figure~6 of \citet{mb09266}. 

Such linear 1-D contours will in general intersect the circle 
described by Equations~(\ref{eqn:circle}) and (\ref{eqn:circle2})
in two
places\footnote{As a special case, the 1-D parallax measurement could
be tangent to the circle (or arc).  It could only miss the circle if
there were systematic errors in either the Earth-based or space-based
data that compromised the result.}.  
Hence, in the general case, the two intersection
points will yield different values of $\pi_\e$, with the fractional
difference being greater when the 1-D contours are farther from being
tangent to the circle.  However, if the parallax circle has been
broken into sufficiently small arclets, then this two-fold discrete
degeneracy may be automatically broken by inconsistency at one
of the two intersection points.

I note that confusion with xallarap effects due to orbital motion of the source
is a potentially more serious problem in the interpretation of
1-D annual parallax compared to 2-D.  (Xallarap has no direct effect
on space-based parallaxes, but is relevant here because I am investigating
1-D annual parallax as a means to break the space-based parallax 
degeneracy.)  Xallarap can, in principle, always perfectly mimic annual
parallax.  However, as pointed out by \citet{poindexter05} it is 
extremely unlikely that, for 2-D parallax, the three principal
xallarap parameters (period, phase, and inclination) would all precisely
mimic those induced by Earth's motion.  But in the case of a putative
1-D annual parallax signal, there is no such strong test against xallarap:
any method of producing uniform acceleration for the main duration of the
event will have exactly the same effect on the light curve.  It is still
the case that xallarap is a priori much less likely than parallax because
the Sun is definitely accelerating Earth in its direction, while only
a small fraction of source stars have companions in the mass and separation
range where they could induce acceleration that is both uniform
(i.e., with sufficiently large semi-major axis) and of sufficient 
strength (i.e., with sufficiently small semi-major axis and sufficiently
large mass) to produce the observed effect.  Nevertheless, this
possibility should be evaluated concretely in each individual case.

\subsection{{Combining with Independent Proper-Motion Information}
\label{lab:pminfo}}

The direction of $\bpi_\e$ is by definition the same as the direction
of $\bmu_\geo$, i.e., the lens-source relative proper motion in the geocentric
frame.  Therefore, if this direction is known, then even the full
circular degeneracy from Equation~(\ref{eqn:circle2}) can be unambiguously
resolved.

There are three known methods to independently measure the lens-source
relative proper motion.  One of these directly measures $\bmu_\geo$,
a second directly measures the heliocentric proper motion
$\bmu_\hel = \bmu_\geo + \bv_\oplus(\pi_\rel/\au)$, while a third directly
measures something that is intermediate.  The relationship between
$\bmu_\geo$ and $\bmu_\hel$ 
has been analyzed in detail by \citet{ob03175} and by \citet{gould14},
and there are no further issues to be explored here.
I mention this issue only for completeness.

Note that because $\theta_\e = \mu_\geo t_\e$ and $t_\e$ is measured
during the event, each of these methods also yields $\theta_\e$, which
is the other parameter (in addition to $\pi_\e$) that is required
to measure $M$ and $\pi_\rel$.

\subsubsection{{Proper Motion From Astrometric Microlensing}
\label{lab:astrometric}}

The light centroid of the two magnified images is displaced from
the true position of the source by,
\begin{equation}
\delta\btheta = -{\Delta\btheta\over (\Delta\theta/\theta_\e)^2 + 2},
\label{eqn:astrometric}
\end{equation}
where $\Delta\btheta$ is the displacement of the lens relative to the
source \citep{my95,hnp95,walker95}.  Thus, by a series of astrometric
measurements (and initially excluding those near the microlensing
event) one can solve for the source parallax $\pi_s$ and proper motion
$\bmu_s$.  Then one can apply Equation~(\ref{eqn:astrometric}) 
to the deviations $\Delta\btheta$ from this solution to determine
$\theta_\e$ and the lens-source relative proper motion.  In practice,
one would fit all the astrometric data to all of these parameters
simultaneously.

Note that if the event is relatively short, then the astrometric
deviations occur while Earth's motion is similar to that at $t_{0,\oplus}$,
so it is the geocentric proper motion that is most directly measured.
If the event is long, then the measurements are most sensitive
to the heliocentric proper motion.  In practice, there is no
ambiguity.  One just, for example, fits for the heliocentric proper
motion and that quantity will be returned by the fitting program.
The distinction is just that if the event is short, the error bars
on a fit to $\bmu_\geo$ will be smaller than on $\bmu_\hel$.

\subsubsection{{Proper Motion By Resolving the Einstein Ring}
\label{lab:ring}}

With sufficiently high resolution, the two images of the source
can be resolved.
In this case, the separation between the
two images and their flux ratio directly yields $\theta_\e$,
while their orientation (position angle $\psi$)
on the sky gives the direction of the
instantaneous lens-source separation $\Delta\btheta$.  The first such image
resolution was recently achieved by \citet{dong19} using VLTI
GRAVITY.

While the direction of lens-source separation $\Delta\btheta$ does not
directly give the direction of lens-source relative proper motion $\bmu_\geo$,
the angle between these two vectors is precisely known from the
photometric light curve, or from the flux ratio of the two images.  
Unfortunately the sign of this angle
(same as the sign of $u_{0,\oplus}$) is not known, and this degeneracy
remains even for the case that we are still considering, $|u_{0,\oplus}|\ll 1$.
In principle, this discrete degeneracy can be resolved by a second
epoch of high-resolution imaging, e.g., 1 day later\footnote{More
precisely, the position angle $\psi$ changes by
$\Delta\psi\rightarrow (\delta t/t_\eff)/[1 + (t-t_0)^2/t_\eff^2]$,
where $t_\eff\equiv u_0 t_\e$ is the effective timescale and $\delta t$
is the elapsed time between the two observations.  This must be significantly
larger than the measurement error of $\psi$.  For the event observed
by \citet{dong19}, $\sigma(\psi) = 0.005$, but other cases may be less
favorable.}.  However, 
\citet{dong19} were unable to obtain a second epoch due to weather.
In such cases, this degeneracy may be resolved by either of the two
methods mentioned above, i.e., by astrometric microlensing or by
1-D annual parallax.  In those cases, either method would itself
give a measurement of the proper-motion direction, but direct imaging
of the Einstein ring gives vastly more precise results.  Hence, the main
role of these auxiliary techniques would simply be to break the degeneracy
\citep{dong19}.  Finally,
this degeneracy could in principle be resolved by the {\it Spitzer}
observations if these restricted the circle to an arc that intersects
one but not both solutions.

\subsubsection{{Proper Motion From Late-Time Imaging}
\label{lab:imaging}}

Finally, after the lens and source separate sufficiently to be separately
resolved \citep{alcock01} or at least to distort their common unresolved
image \citep{bennett06}, then their relative proper motion can be determined 
simply by dividing their measured vector separations by the elapsed
time since $t_{0,\oplus}$.  

In contrast to the previous three methods,
which do not depend in any way on the lens being luminous, this method
appears at first sight to require a luminous lens.  And it therefore
appears to be less valuable, because if the lens can be imaged (which
automatically yields $\bmu$ and so an estimate of $\theta_\e = \mu_\geo t_\e$),
then good estimates of $M$ and $\pi_\rel$ 
can already be made from its photometric properties
combined with the constraint $\theta_\e^2=\kappa M\pi_\rel$.  However,
as discussed by \citet{gould14}, this ``lower value'' is somewhat
deceptive.  For several tens of percent of cases, the star that is imaged
will actually be a binary companion to a dimmer (or possibly dark) lens.
The proper motion of this companion will be nearly identical to
that of the lens, but its photometric properties will be completely
unrelated to those of the lens.  Such cases can only be detected and
analyzed if there is an independent measurement of the microlens parallax.
Because the proper motion is measured by this imaging, all that is
required to extract a full 2-D determination of $\bpi_\e$ is a 1-D
parallax.  This could come from 1-D annual parallax \citep{ob03175,gould14},
or from a circle (or arc) as investigated in the current work.

\section{{General Case of a Single Late-Time Observation}
\label{sec:general}}

\begin{figure}
\centering
\includegraphics[width=90mm]{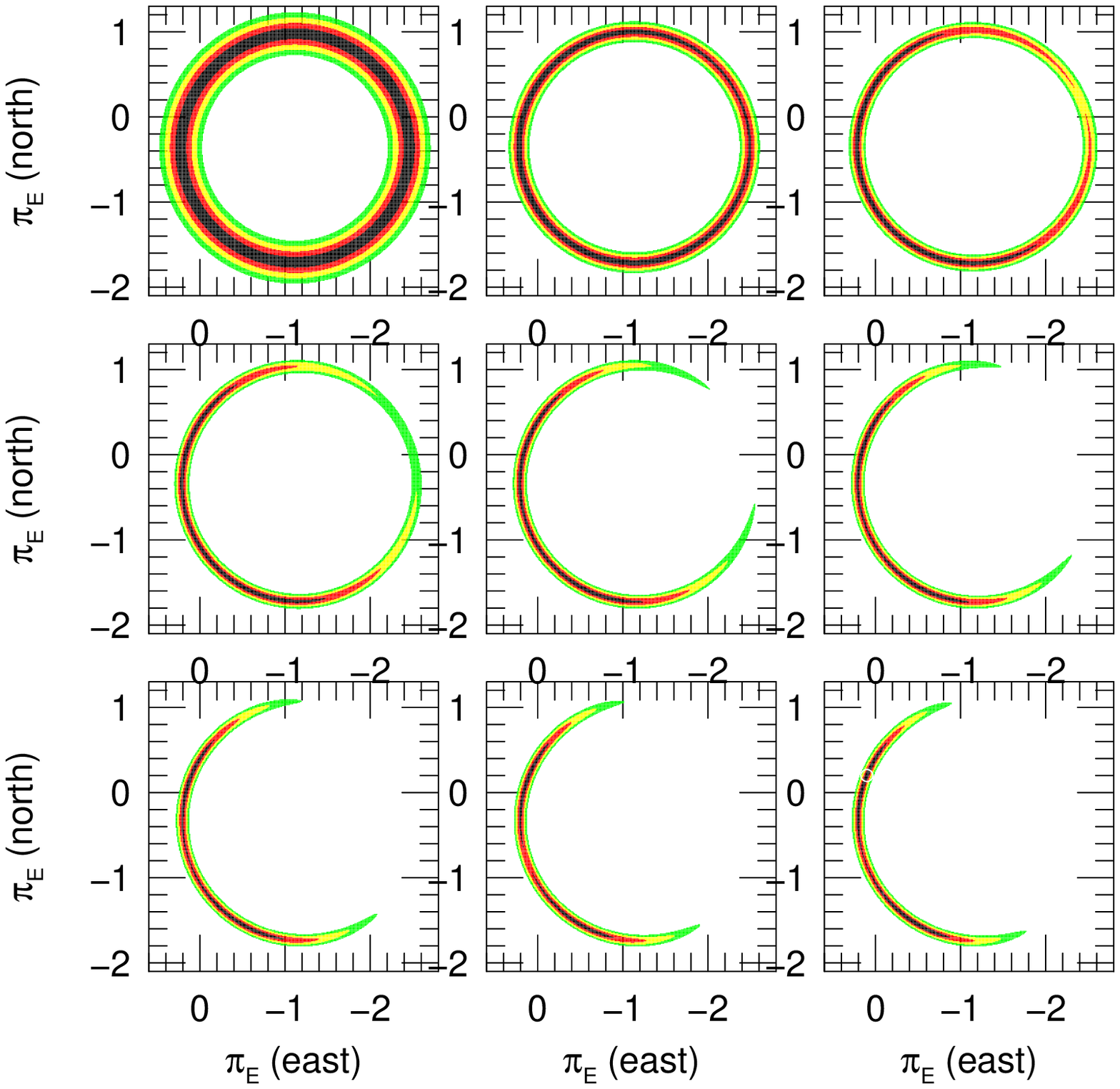}
\caption{Similar to Figure~\ref{fig:all_0} except that the event
as seen from Earth has $u_{0,\oplus}=+0.4$ (rather than 0).  Note that
the symmetry axis is inclined to the $\bQ$ axis (basically east-west)
by $\simeq\tan^{-1}(u_0/\Delta\tau)$, i.e., the same as the angle between
$\bP_\e$ and $\bpi_\e$ (see Eq.~(\ref{eqn:bpdef})).  This is contrary to
the case that the satellite observations cover the peak 
\citep{refsdal66,gould94}.
Because the offset between the two degenerate solutions is larger 
($|\bpi_{\e,++} - \bpi_{\e,+-}|\sim 1.6$ versus $\sim 0.4$ for $u_{0,\oplus}=0$
from Fig.~\ref{fig:all_0}), the large arc tends to break up into
two arclets, although only marginally for the adopted measurement errors
$\sigma=0.01\,$mag shown here.
}
\label{fig:all_u0p}
\end{figure}

I began by investigating the special case of high-magnification 
($u_{0,\oplus}$) because it captures the essential physics and is
mathematically simple.  But it is important to also explore the
more general case.  To facilitate this investigation, I introduce
$\bpi_\e^T$, which I define as having the same magnitude as $\bpi_\e$
($|\bpi_\e^T|=\pi_\e$), but whose direction is orthogonal
($\bpi_\e^T\cdot \bpi_\e=0$).  And I introduce another vector
\begin{equation}
\bP_\e \equiv {(\Delta\tau)\bpi_\e + u_0\bpi_\e^T\over
\sqrt{(\Delta\tau)^2 + u_0^2}},
\label{eqn:bpdef}
\end{equation}
which also has the same magnitude ($|\bP_\e|=\pi_\e$) but is rotated
relative to $\bpi_\e$ by $\tan^{-1}(u_0/\Delta\tau)$.  Note that
I have suppressed the ``$\oplus$'' subscript on $u_0$.
Then, Equation~(\ref{eqn:usat1}) becomes
\begin{equation}
u_\sat^2= \bigg|{\bpi_\e\Delta\tau\over \pi_\e}-\bQ\pi_\e + 
{\bpi_\e^T\over\pi_\e}u_{0}\bigg|^2
\label{eqn:u2sat1}
\end{equation}
or
\begin{equation}
u_\sat^2 = (\Delta\tau)^2 +u_0^2 -2\bQ\cdot\bP_\e\sqrt{(\Delta\tau)^2+u_0^2} + 
Q^2P_\e^2 .
\label{eqn:u2sat2}
\end{equation}
Similarly to Equation~(\ref{eqn:usat2}), this can be rewritten as
\begin{equation}
u_\sat^2 = \bigg|Q\bP_\e
- {\bQ\over Q}\sqrt{(\Delta\tau)^2 + u_0^2}\bigg|^2
\label{eqn:u3sat2}
\end{equation}
or
\begin{equation}
\bigg|\bP_E - {\bQ\over Q^2}\sqrt{(\Delta\tau)^2 + u_0^2}\bigg|^2 
= \biggl({u_\sat\over Q}\biggr)^2.
\label{eqn:u2sat3}
\end{equation}
That is, formally, $\bP_\e$ traces a circle with center
$\sqrt{(\Delta\tau)^2+u_0^2}\bQ/Q^2$ and radius $u_\sat/Q$,
\begin{equation}
\bP_\e =\sqrt{(\Delta\tau)^2 + u_0^2}{\bQ\over Q^2} + {u_\sat\over Q}\hat{\bf n}.
\label{eqn:circle3}
\end{equation}
I now express this vector equation in a specific coordinate system,
in which the $x$-axis is aligned with $\bQ$ and the $y$-axis is orthogonal
to it.  The center of the $\bP_\e$ circle is then at
$(\sqrt{(\Delta\tau)^2 + u_0^2},0)$.  Because $\bP_\e$ and $\bpi_\e$
are related by a simple rotation of $\pm\tan^{-1}(u_0/\Delta\tau)$
(depending on the sign of $u_0$) the contour for $\bpi_\e$ will still
be a circle of the same radius, but with its center rotated by this angle.
That is, in this same coordinate system,
\begin{equation}
\bpi_{\e,\pm}(\phi) = {(\Delta\tau,\pm|u_0|) + (\cos\phi,\sin\phi)u_\sat\over Q},
\label{eqn:circle4}
\end{equation}
where $\phi$ parameterizes the position around the circle, and the 
``$\pm$'' subscript
shows the solutions for the two different signs\footnote{To be consistent
with the generally used sign convention that is described in Figure~4
of \citet{gould04}, the center of the parallax circle in 
Equation~(\ref{eqn:circle4}) should be expressed as
$[(Q_N\Delta\tau + Q_E u_0),(Q_E\Delta\tau - Q_N u_0)]/Q^2$.
}
 of $u_0$ (i.e., $u_{0,\oplus}$).
That is, there are two circles of the same size, whose centers are offset
by $\pm u_0/Q$ in the direction orthogonal to $\bQ$.

Figure~\ref{fig:all_u0p} shows the results of the same observation
sequence as Figure~\ref{fig:all_0} but assuming that the otherwise
identical event has $u_{0,\oplus}=+0.4$.  Note that axis of symmetry is
inclined to the $\bQ$ axis (essentially the east-west axis) by about
$\tan^{-1}(u_\sat/\Delta\tau)$.  This is contrary to usual case,
which was analyzed by \citet{refsdal66} and \citet{gould94}, 
for which the symmetry axis is along the Earth-satellite separation vector.
Also note that in this case, the single arc is beginning to break
up into arclets, one centered on each of the two degenerate solutions.

\section{{Discussion}
\label{sec:discuss}}

While I began my investigation of circular microlens-parallax degeneracies 
with the specific aim of understanding the $\bpi_\e$ ``arcs'' that appear
in microlensing events with late-time {\it Spitzer} observations,
these circular degeneracies are actually a powerful tool for understanding
space-based microlensing parallaxes more generally.  In fact, as mentioned
in Section~\ref{sec:ideal}, Equations~(\ref{eqn:circle}), (\ref{eqn:circle2}), 
(\ref{eqn:circle3}), and (\ref{eqn:circle4}) 
actually apply to {\it any} individual space-based
observation (provided that $F_{s,\sat}$ and $F_{s,\rm base}$ are known).

\subsection{{Parallax Circles: A General Tool}
\label{sec:generaltool}}

That is, any ensemble of satellite microlensing parallax observations
can be understood as an overlapping set of circles on the $\bpi_\e$ plane.
In Section~\ref{sec:impact}, I gave one application of this approach
to understand how these overlapping circles combine to form arcs
for events with only late-time satellite observations.

\begin{figure}
\centering
\includegraphics[width=90mm]{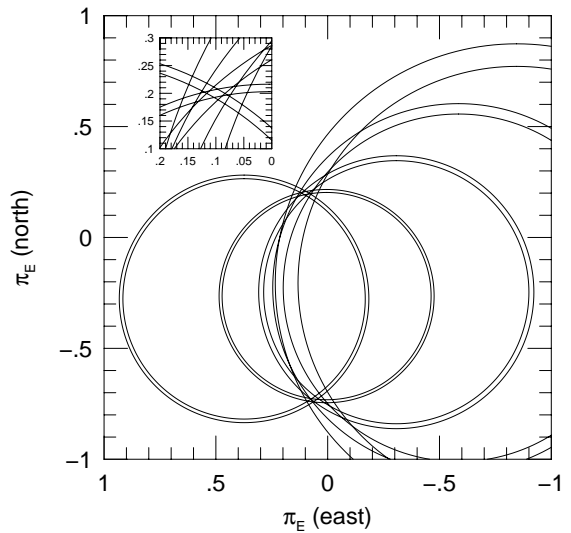}
\caption{Five single-measurement circular degeneracy contours 
in the $\bpi_\e$ plane shown
for an event with $u_{0,\oplus}=+0.4$, $t_\e=18\,$days and 
$t_{0,\oplus}=$18 July 2019.
The five simulated {\it Spitzer} measurements are (left to right) at 
$\Delta\tau=(t-t_{0,\oplus})/t_\e = (-0.5,0,+0.5,+1.0,+1.5)$.  
The spacing of the concentric circles shows $\pm 1\,\sigma$ errors propagated
from measurement errors of $\pm 0.01\,$mag.
Any combination
of two of the first three observations would give a very precise
measurement of $\bpi_\e$ (up to a two-fold degeneracy).  However,
because one does not know a priori how the space and ground events are
offset, roughly four measurements would be necessary to reasonably
guarantee good precision (plus an additional measurement at baseline).
See \citet{calchi16} for a view of the parallax geometry in the
heliocentric-observer (as opposed to $\bpi_\e$) plane.
}
\label{fig:cc}
\end{figure}

\subsection{{Understanding the Four-Fold Degeneracy}
\label{sec:fourfold}}

A second application is to provide an alternate understanding of the
four-fold degeneracy.  Let us first consider the case of $u_{0,\oplus}=0$,
for which there is a two-fold degeneracy.  The circles are always
centered on the $\bQ$ axis.  They all must pass through the actual
value of $\bpi_\e$.  In doing so, they must also pass through
$\bpi_\e^\prime = 2[(\bQ/Q)\cdot\bpi_\e](\bQ/Q) - \bpi_\e$, which is
as much ``below'' the $\bQ$ axis as $\bpi_\e$ is above it.  Because 
this expression depends only on the direction of $\bQ$
and not its magnitude, all circles that pass through $\bpi_\e$ will
also pass through $\bpi_\e^\prime$ provided that this direction does not change.
Hence, breaking this degeneracy (from satellite data alone) depends
on $\bQ$ changing direction enough to have a significant effect.

For the case of $u_0\not=0$, the picture of intersecting circles
while the source is within the Einstein ring directly reproduces
the traditional understanding of the two-fold degeneracy between
the source passing on the same versus opposite sides of the lens
as seen from the two observatories.  See Figure~\ref{fig:cc}.  However,
as shown by Figure~\ref{fig:all_u0p}, the symmetry axis
rotates at late times, thus providing some possibility that the late-time
``arc'' will be inconsistent with one of the two solutions.

\subsection{{Cheap Satellite Parallaxes at All Magnifications}
\label{sec:cheap}}

The geometry shown in Figure~\ref{fig:cc}
immediately gives rise to a third and fourth application.
The third application is a generalization of the \citet{gouldyee12}
proposal for ``cheap space-based microlens parallaxes''.  Recall that
their proposal rested on obtaining a space-based image very near 
$t_{0,\oplus}$ and was restricted to high-magnification events $u_{0,\oplus}\ll 0$.
However, using the circle picture, it is easy to see that two 
satellite observations (plus baseline) are all that are needed 
in principle to measure $\bpi_\e$.  That is, two circles, regardless
of relative size, can only intersect in zero, one, or two places.
Parallax circles must intersect at least once (within errors) at $\bpi_\e$.

If the circles intersect in two places (i.e., cross rather than being
tangent), then the $\Delta\chi^2=2$ error contour (containing
$(1-e^{-1})=63\%$ of the probability) is given directly
by the ellipse that passes through the four intersection points of the
two sets of $1\,\sigma$ error-circles.  Consider, for example, 
the two smallest-error circles in Figure~\ref{fig:cc}.  One sees 
from the inset that these intersection points are separated
by $\Delta\pi_{\e,N}=0.014$ along the ordinate and 
by $\Delta\pi_{\e,E}=0.044$ along the abscissa.  Hence
$\sigma(\pi_{\e,N},\pi_{\e,E})
= (\Delta\pi_{\e,N},\Delta\pi_{\e,E})/\sqrt{8} = (0.005,0.016)$.

While it is always possible ``in principle'' to determine $N$ parameters
(in the case $N=2$) from $N$ measurements, there are two main
practical issues that usually lead one to seek some redundancy, i.e.,
more data points.  First, some measurements may turn out to be
mathematically degenerate (or nearly degenerate).  Second, usually
one would like to have internal checks on the externally calibrated
error bars.  I address these issues in turn.

If the two circles are tangent (or nearly so), and therefore have effectively
only one point of intersection, then (after taking account of measurement
errors), their overlap will be an arc.  I have already shown 
in Section~\ref{sec:impact} that
such arcs are the natural consequence of a late-time series of
satellite observations.  The main way to avoid osculating circles 
(and so arc or even circle degeneracies) is to make the observations
while the source is inside the Einstein ring as seen from the satellite.
This is easier said than done because one does not know a priori when
this will occur.  Indeed, for very large $\pi_\e\gtrsim 1$, there is
no guarantee that the source will even pass within the Einstein ring
as seen from the satellite.  Thus there is some chance that one or both
of two well-chosen observations, e.g., at $t_{0,\oplus}\pm 0.5\,t_\e$
would fall outside the Einstein ring.  Hence, a more aggressive approach
would be to make the first two observation early, e.g., at 
$t_{0,\oplus}- 0.5\,t_\e$ and $t_{0,\oplus}- 0.3\,t_\e$, to determine whether
the event was rising or falling and make a rough $\bpi_\e$ measurement.  And
then to use these to decide on one or two additional measurements (in addition
to baseline).  Still, only of order $4+1=5$ observations would be needed.
While more than the absolute minimum of $2+1=3$, it is still far
less than in the current mode.  See Figure~\ref{fig:cc}.

Note that this approach could not be applied to {\it Spitzer}
microlensing for three reasons.  First, the observations are initiated 
with a 3-10 day delay.  See Figure~1 of \citet{ob140124}.  This
means that the great majority of events have their first observation
after $t_{0,\oplus}- 0.5\,t_\e$, which very often proves to be near or after
$t_{0,\sat} + 0.5\,t_\e$.  Second, there is no way to alter the
observing schedule on a daily basis as envisaged in the previous paragraph.
Third, the data are not downloaded fast enough to make such real time
decisions.  However, if a satellite were specifically engineered for microlens
parallaxes, then it could incorporate these capabilities.  This would
make it possible to monitor 1500 microlensing events per year with
only 30 observations per day, which implies that a very small telescope
(e.g., 20 cm) would be adequate.

A second reason for obtaining $4+1$ (rather than $2+1$) observations
is to control systematics, i.e., deviations of the measured
versus true values that are not captured by the statistical error bars.
These can take a variety of forms, but the two of greatest concern
are large random fluctuation (due to stochastic processes on either
the sky, e.g., cosmic ray events, or the detector) and long-term
trends.  Of course, any satellite undergoes extensive commissioning
observations at the start of the mission that are matched to its 
envisaged scientific goals, which would characterize such systematics
in the present case.  Still, it would be useful to have ongoing checks
against large stochastic outliers, which would be a routine by-product
of $(4+1)$ observations.  (It is likely that the $\sim 1\,$hr exposures
mentioned above would be subdivided into several sub-exposures, which
would provide additional redundancy.)

The problem of long term trends (so-called ``red-noise'') is substantially
less severe when parameters are derived from a few measurements 
(the present case) as opposed to many measurements (e.g., transiting planets).
To understand this concretely, consider the {\it Spitzer} light curve
of OGLE-2016-BLG-1045, in Figure~1 of \citet{ob161045}.  This shows
long term residual trends with semi-amplitude $\sim 0.015\,$mag,
which is approximately equal to the statistical errors.  Suppose
that one searched for a planetary transit in a region of a light curve
containing $N=400$ points with these statistical and systematic error 
properties, and derived a transit with the same depth, i.e.,
0.015 mag. If one treated the errors as being purely statistical
at $\sigma=0.015\,$mag, then the error in the transit depth would
be $\sigma/\sqrt{400} = 0.00075\,$mag, so a clear $\Delta\chi^2=400$
planet ``detection''.  Even if one added the systematic ``noise'' 
$0.015/\sqrt{2}$ in quadrature to the statistical noise, one would 
still end up with a spurious $\Delta\chi^2=267$ ``detection''
(if one continued to treat the 400 individual measurements as independent).

But note that no such  issue of ``red'' (i.e., correlated) noise arises
in the single-epoch measurement tested by \citet{ob161045} for
OGLE-2016-BLG-1045.
The one measurement that they use has an empirically renormalized
error bar that automatically takes account of deviations due
to both long term trends and statistical fluctuations. It does
not take account of correlations, but since there is only
a single measurement, correlations do not play any role.

The situation is only slightly worse for the two-measurement 
determinations envisaged here.  That is, if the error bars
were set to account for both random fluctuations and systematic
trends (as in case of OGLE-2016-BLG-1045), then the pairs of
error circles in Figure~\ref{fig:cc} would each individually
be correct.  It would not be correct to treat two measurements
as statistically independent, but if one did so nevertheless
(and if, e.g., the amplitude of systematic trends were equal
to the statistical fluctuations as in OGLE-2016-BLG-1045), 
then one would only underestimate the true error of the $\bpi_\e$
determination by factor $\sqrt{6/5}\sim 1.1$.  Of course, one
should not proceed in this naive way, but rather take proper
account of the actual error properties of the data.  Nevertheless
this exercise shows that ``red noise'' is a relatively minor
issue for parameter measurements based on a few measurements
unless the red noise itself is very severe.  For the case of {\it
Spitzer}, \citet{zhu17} found severe red noise in only a small 
fraction of the order 50 events with data adequate to make this
determination.  The origin of these systematics is not precisely
known, but is unlikely to affect an optical satellite with
subsampled pixels of near uniform response, which (in contrast
to {\it Spitzer}) would permit standard difference imaging analysis
(DIA, \citealt{alard98}).


\begin{figure}
\centering
\includegraphics[width=90mm]{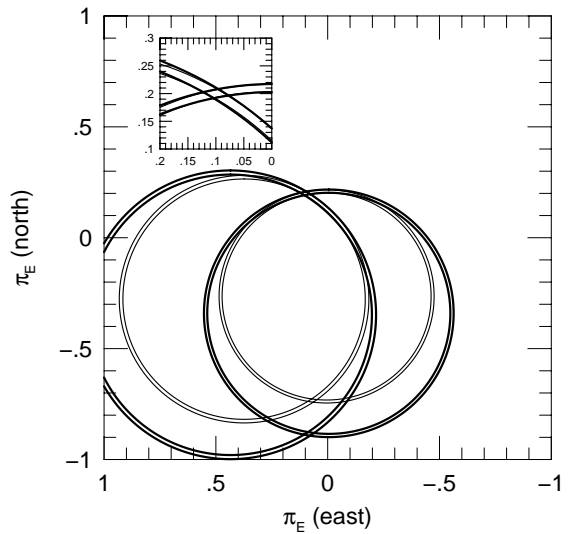}
\caption{Illustration of degeneracy breaking using two satellites.
The two pairs of thin-lined concentric circles are exactly the same as
the parallax circles in Figure~\ref{fig:cc} that are closest to peak,
namely at $\Delta\tau=(t-t_{0,\oplus})/t_\e = -0.5$ and $+0.5$ and for
exactly the same simulated event.  The intersect at the true parallax
$(\pi_{\e,N},\pi_{\e,E})=(0.2,0.1)$, but also at a second degenerate
solution, $(\pi_{\e,N},\pi_{\e,E})\simeq(-0.7,0.1)$.  Indeed parallax
circles at all epochs intersect at approximately the same locations.
See Figure~\ref{fig:cc}.  The thick-lined concentric circles correspond
to measurements made by a hypothetical second satellite in a {\it Spitzer}
like orbit but launched exactly six years later and so not as far
from Earth.  These circles also intersect in two degenerate location,
i.e., $(\pi_{\e,N},\pi_{\e,E})=(0.2,0.1)$ and $(-0.9,0.1)$.  Combining
the measurements of the two satellites resolves the degeneracy
in favor of the first solution.
}
\label{fig:cc2}
\end{figure}

\subsection{{Cheap Breaking of the Four-Fold Degeneracy}
\label{sec:break}}

The simplified approach outlined above would still leave the four-fold
degeneracy in tact. In many cases this could be broken by one of the four
methods outlined in Section~\ref{sec:resolution}.  However, this
simplified approach also makes it feasible to carry out the monitoring
with two such satellites, as originally envisaged by \citet{refsdal66}.
If both satellites were near the ecliptic (by far the cheapest approach), 
then only one pair of degeneracies would generally be broken.  This was the
outcome when \citet{mb16290} applied this two-satellite technique, using
{\it Spitzer} and {\it Kepler}, both of which are near the ecliptic.
However, the degeneracy that was broken (between 
$\Delta\beta^\prime_{\pm,\mp}$ and $\Delta\beta^\prime_{\pm,\pm}$, i.e,
opposite versus same signs) is by far the more important one because
it would lead to different magnitudes of $\pi_\e$ and so different
lens masses $M=\theta_\e/\kappa\pi_\e$ and lens-source relative parallaxes
$\pi_\rel = \theta_\e\pi_\e$. Thus, a second, small, low-cost satellite 
would be by far the simplest and most
robust method to systematically remove this degeneracy.  
See Figure~\ref{fig:cc2}



\acknowledgments

I thank Subo Dong for both fruitful discussions and critiquing the manuscript.
I thank the anonymous referee, whose comments and suggestions 
significantly improved the manuscript.
This work was supported by NSF grant AST-1516842 and by JPL grant 1500811.
I received support from the European  Research  Council  under  the  European  Union’s Seventh Framework Programme (FP 7) ERC Grant Agreement n. [321035].

\end{document}